\documentclass[conference]{IEEEtran}

\usepackage{cite}
\usepackage{amsmath,amssymb,amsfonts}
\usepackage{algorithm}
\usepackage{array}
\usepackage{graphicx}
\usepackage{textcomp}
\usepackage{xcolor}
\usepackage{stfloats}
\usepackage{verbatim}
\usepackage{cite}

\usepackage[a4paper, total={184mm,239mm}]{geometry}
\def\BibTeX{{\rm B\kern-.05em{\sc i\kern-.025em b}\kern-.08em
    T\kern-.1667em\lower.7ex\hbox{E}\kern-.125emX}}

\usepackage[caption=false]{subfig}
\usepackage{url}
\usepackage{color}
\usepackage{algpseudocode}
\usepackage{algorithmicx}

\newcommand{\lfscale}{.53}

\algnewcommand\algorithmicforeach{\textbf{for each}}
\algdef{S}[FOR]{ForEach}[1]{\algorithmicforeach\ #1\ \algorithmicdo}

\usepackage{ifthen}
\usepackage[smaller,nolist,nohyperlinks]{acronym}

\usepackage{listings}

\lstdefinelanguage{LF}{
  keywords={target, deadline, after, state, logical, physical, startup, shutdown, reaction, preamble, x, reactor, trigger, input, output, constructor, new, action, clock, actor, handler, int, main, Main, timer, sec, secs, msec, msecs, usec, usecs, mode, initial, reset, continue, history, if, else},
  keywordstyle=\color{black}\bfseries,
  ndkeywords={class, export, boolean, throw, implements, import, this},
  ndkeywordstyle=\color{darkgray}\bfseries,
  identifierstyle=\color{black},
  sensitive=false,
  comment=[l]{//},
  morecomment=[s]{/*}{*/},
  commentstyle=\color{purple}\ttfamily,
  stringstyle=\color{black}\ttfamily,
  morestring=[b]',
  morestring=[b]"
}

\newcommand{\code}[1]{\texttt{\small #1}}

\newlength{\linenumwidth} 
\makeatletter
\def\lst@PlaceNumber{%
  \makebox[\linenumwidth][l]{\normalfont\lst@numberstyle{\thelstnumber}}%
}
\makeatother

\usepackage{xspace}
\usepackage[scaled=0.88]{helvet}

\newcommand{\eg}{e.\,g.\xspace}

\newcommand{\etal}{et\,al.\xspace}
\newcommand{\wrt}{w.\,r.\,t.\xspace}

\usepackage[bookmarks=false]{hyperref}

\begin{document}

\begin{acronym}
  \acro{cfg}[CFG]{Controlflow Graph}
  \acro{fsm}[FSM]{finite-state machine}
  \acro{ide}[IDE]{integrated development environment}
  \acro{kieler}[KIELER]{KIEL Integrated Environment for Layout Eclipse Rich Client}
  \acro{led}[LED]{light-emitting diode}
  \acro{lf}[LF]{Lingua Franca}
  \acro{scade}[SCADE]{Safety-Critical Application Development Environment}
  \acro{uml}[UML]{Unified Modeling Language}
\end{acronym}
\acused{ide}
\acused{uml}
\acused{led}
\acused{scade}

\title{Modal Reactors}

\author{%
  \IEEEauthorblockN{%
    Alexander Schulz-Rosengarten\IEEEauthorrefmark{1},
    Reinhard~von~Hanxleden\IEEEauthorrefmark{1},
    Marten Lohstroh\IEEEauthorrefmark{2},
    Soroush Bateni\IEEEauthorrefmark{3},
    Edward A. Lee\IEEEauthorrefmark{2}
    \IEEEauthorblockA{\IEEEauthorrefmark{1}
      \textit{Kiel University}, Kiel, Germany, \{als, rvh\}@informatik.uni-kiel.de}
    \IEEEauthorblockA{\IEEEauthorrefmark{2}
      \textit{UC Berkeley}, California, USA, \{marten, eal\}@berkeley.edu}
    \IEEEauthorblockA{\IEEEauthorrefmark{3}
      \textit{UT Dallas}, Texas, USA, soroush@utdallas.edu}
  }
}

\maketitle

\begin{abstract}

Complex software systems often feature distinct modes of operation, each designed to handle a particular scenario that may require the system to respond in a certain way.
Breaking down system behavior into mutually exclusive modes and discrete transitions between modes is a commonly used strategy to reduce implementation complexity and promote code readability.

However, such capabilities often come in the form of self-contained domain specific languages or language-specific frameworks.
The work in this paper aims to bring the advantages of modal models to mainstream programming languages, by following the polyglot coordination approach of \ac{lf}, in which verbatim target code (\eg, C, C++, Python, Typescript, or Rust) is encapsulated in composable reactive components called reactors.
Reactors can form a dataflow network, are triggered by timed as well as sporadic events, execute concurrently, and can be distributed across nodes on a network.

With modal models in \ac{lf}, we introduce a lean extension to the concept of reactors that enables the coordination of reactive tasks based on modes of operation.
The implementation of modal reactors outlined in this paper generalizes to any \ac{lf}-supported language with only modest modifications to the generic runtime system. 
\end{abstract}


\section{Introduction}
\label{sec:intro}

The focus of this paper is on reactive systems, which continuously react to their environment, are typically embedded in larger systems, and often have some real-time requirements.

Two major notations or views have emerged for describing reactive systems, actor-oriented dataflow networks and state machines.
The \emph{dataflow view} breaks down the program into smaller blocks 
with streams of data flowing between them.
Each such actor receives inputs, produces outputs, and can be assumed to operate fully independently from other blocks on which it has no data dependencies, thereby presenting opportunities for parallelization or distribution.
MathWorks' Simulink and National Instruments' LabVIEW are examples for such an approach.

In a \emph{state-oriented view}, the program is modeled in terms of states of the system and its progression in the form of transitions between them.
State machine notations can be found, for example, as Stateflow~\cite{hamon2004operational} in Simulink or as Statecharts~\cite{harel1987statecharts}.
While state machines often describe fine-grained steps at the system level, they can also be used to represent more abstract \emph{modes} of operation.
For example, a system or subsystem may progress from initialization mode, through a training mode, and into a steady-state mode, with additional modes for error handling.
Each mode of operation may encapsulate a complex collection of (stateful) reactive behaviors.
Such \emph{modal models} were realized, for example, in Ptolemy II~\cite{LeeTripakis:10:ModalModels}, where they were used to simulate complex and hybrid systems.

However, the languages that provide the capabilities to model systems in any of these notations often come in the form of standalone domain specific languages (as in Simulink, LabVIEW, or Ptolemy II) or language-specific frameworks (such as Akka~\cite{AkkaAction2016}).
Usually, these languages either compile to or integrate into specific general purpose programming languages to produce executable code.
The idea of \emph{polyglot coordination} is to allow any mainstream programming languages to benefit from the advantages of modeling with actors, states, or modes.
This can be done by directly embedding the verbatim code and then producing executable code that coordinates the execution of these modular units.

The goal of the work in this paper is to bring the advantages of modal models
to mainstream programming languages through a reactor-oriented coordination language
called Lingua Franca (\ac{lf})~\cite{LohstrohEtAl:21:Towards}.
\ac{lf} is rooted in a model of computation called \emph{reactors}~\cite{Lohstroh:2019:CyPhy} and is built as a polyglot coordination language.
Reactors encapsulate reactive tasks specified in verbatim code and provide a \emph{minimal coordination layer} around them that is reactive, timed, concurrent, event-based, and accounts for isolated states.
Unlike Ptolemy II, \ac{lf} is not merely intended for modeling and simulation, but rather is meant for building efficient implementations.
\ac{lf} currently supports C, C++, Python, TypeScript, and Rust.
For these languages it provides a runtime environment for automatic coordination of time-sensitive and concurrent or distributed reactors.
The applicability of \ac{lf} ranges from embedded systems to distributed systems
deployed to the Cloud.

\subsection*{Contributions and Outline} 

While the reactor-oriented modeling approach is well-suited for concurrent and distributed event-based coordination, it does not allow to naturally coordinate these tasks in terms of modes of operation, as we will illustrate in \autoref{sec:furuta}.
After summarizing relevant aspects of \ac{lf} in \autoref{sec:reactors}, we will present our concept of modal reactors, extending the polyglot coordination layer of \ac{lf}.
Specifically, this includes
\begin{itemize}
  \item a lean textual and diagrammatic modal language extension that embraces the polyglot nature of \ac{lf} by taking a ``black-box approach'' towards the target language and that allows hierarchical decomposition of modal behavior; 
  \item an adaptation to reactors with two simple but effective transition types, reset and history, which can be further refined at the target language level; and
  \item a semantics for modal behavior that introduces mode-local time and leverages \ac{lf}'s superdense time model to achieve deterministic behavior.
\end{itemize}
\autoref{sec:implementation} gives a closer look at implementation aspects and \autoref{sec:discussion} elaborates on crucial design decisions and potential alternatives.
We discuss related work in \autoref{sec:related-work} and conclude in \autoref{sec:conclusions-outlook}.

\section{Motivating Example: The Furuta Pendulum}
\label{sec:furuta}

\begin{figure}
  \centering
  \includegraphics[width=.8\linewidth]{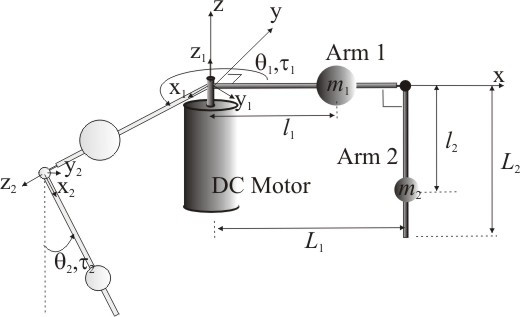}
  \label{fig:furuta-schematic}
  \caption{Schematic of the Furuta pendulum from \href{https://en.wikipedia.org/wiki/Furuta\_pendulum\#/media/File:Furuta\_pendulum.jpg}{Wikipedia} by Benjamin Cazzolato | \href{https://creativecommons.org/licenses/by/3.0/}{CC BY 3.0}.}
\end{figure}

A Furuta pendulum~\cite{furuta1992swing} is a classic control system problem often used to teach feedback control.
As shown in \autoref{fig:furuta-schematic}, it consists of a vertical shaft driven by motor, a fixed arm extending out at 90 degrees from the top of the shaft, and a pendulum at the end of the arm.
The goal is to rotate the shaft to impart enough energy to the pendulum that it swings up,
to then catch the pendulum and balance it so that the pendulum remains above the arm.
Each of these steps requires a different control behavior which makes a controller a prime candidate for a modal model.
It cycles through the three modes, which we will name \code{SwingUp}, \code{Catch}, and \code{Stabilize}.

From a classical event-driven or dataflow perspective, there is only a single reactive task, computing the motor control based on the angle measurements at the arm and shaft.
However, with modes we can identify more fine-granular tasks and coordinate these by embedding them in a modal model.

\begin{figure}
  \centering

  \subfloat[Structural overview as graphical diagram]{
    \centering
    \includegraphics[scale=\lfscale]{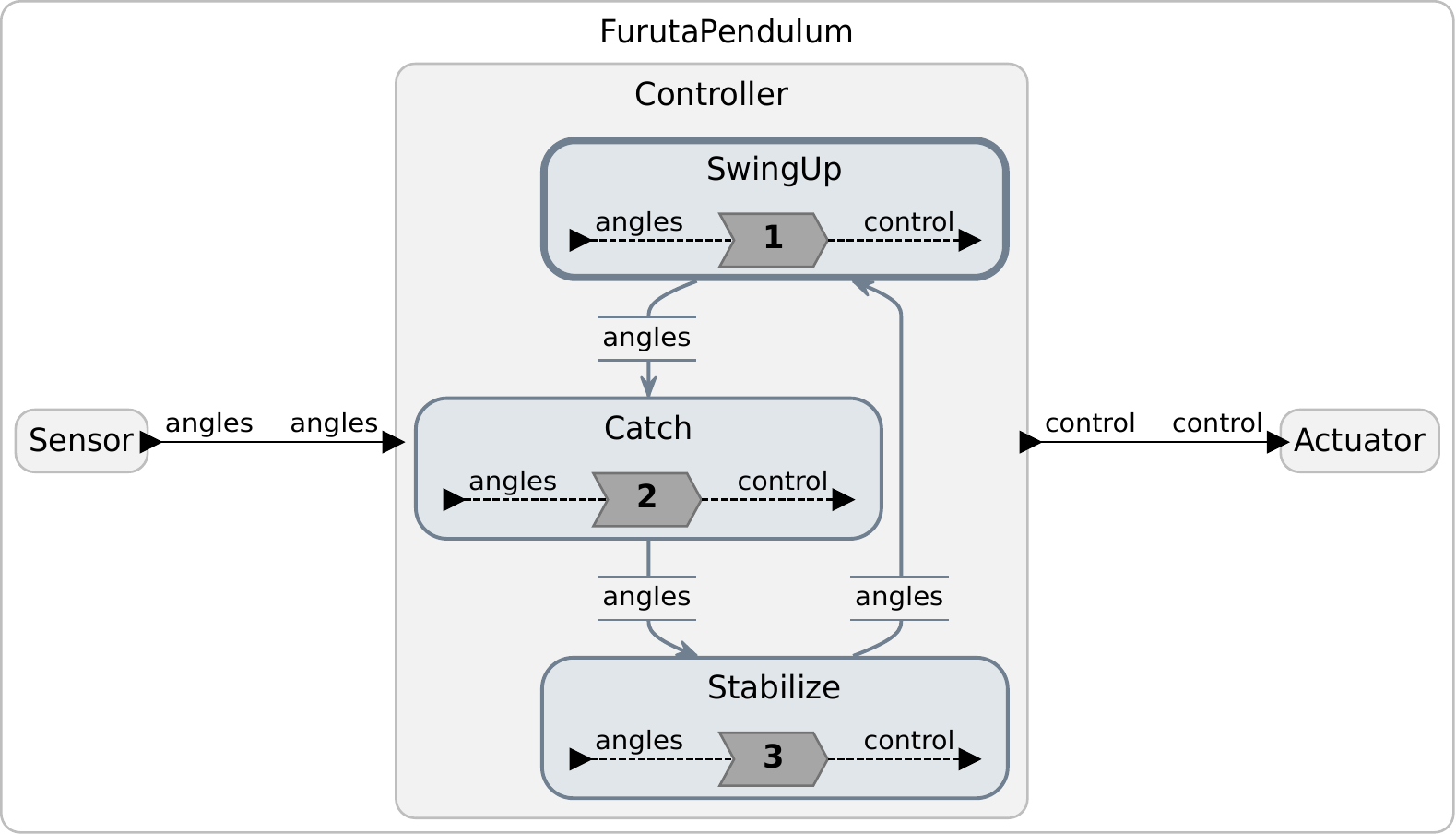}
    \label{fig:furuta-diagram}
  }

  \subfloat[The \code{Controller} reactor code]{
    \begin{minipage}[b]{\linewidth}
    \centering
\begin{lstlisting}[language=LF,escapechar=|]
target C;
reactor Controller {
  input angles:float[];
  output control:float;
  initial mode SwingUp {|\label{ln:swingup}|
    reaction(angles) -> control, reset(Catch) {=|\label{ln:r1}|
      ... control law here in C ...|\label{ln:cl1}|
      lf_set(control, ... control value ... );|\label{ln:out1}|
      if ( ... condition ... ) {|\label{ln:cond1}| lf_set_mode(Catch);|\label{ln:mt1}| }
    =}
  }
  mode Catch {|\label{ln:catch}|
    reaction(angles) -> control, reset(Stabilize) {=|\label{ln:r2}|
      ... control law here in C ...|\label{ln:cl2}|
      lf_set(control, ... control value ... );|\label{ln:out2}|
      if ( ... condition ... ) {|\label{ln:cond2}| lf_set_mode(Stabilize);|\label{ln:mt2}| }
    =}
  }
  mode Stabilize {|\label{ln:stablize}|
    reaction(angles) -> control, reset(SwingUp) {=|\label{ln:r3}|
      ... control law here in C ...|\label{ln:cl3}|
      lf_set(control, ... control value ... );|\label{ln:out3}|
      if ( ... condition ... ) {|\label{ln:cond3}| lf_set_mode(SwingUp);|\label{ln:mt3}| }
    =}
  }
}
\end{lstlisting}
    \label{fig:furuta-code}
    \end{minipage}
  }

  \subfloat[The main reactor code for the program]{
    \begin{minipage}[b]{\linewidth}
    \centering
\begin{lstlisting}[language=LF,escapechar=|]
target C;
import Controller from "FurutaPendulumController.lf"
import Sensor, Actuator from "FurutaPendulumUtil.lf"

main reactor {|\label{ln:main}|
    s = new Sensor();
    c = new Controller();
    a = new Actuator();
    s.angles -> c.angles;|\label{ln:connect1}|
    c.control -> a.control;|\label{ln:connect2}|
}
\end{lstlisting}
    \label{fig:furuta-impl-code}
    \end{minipage}
  }

  \caption{\ac{lf} program to drive the Furuta Pendulum.}
  \label{fig:furuta}
\end{figure}

To illustrate what we are aiming for with modal reactors, we have replicated a solution given by Eker et al.~\cite{liu2002realistic} and implemented it using our mode extension for Lingua Franca.
The program is presented in \autoref{fig:furuta}.
Our language extension includes the diagram synthesis capabilities of \ac{lf}, which yields an automatically generated and interactive pictorial representation (\autoref{fig:furuta-diagram}) of the textual program.
The overall program consists of three connected reactors \code{Sensor}, \code{Controller}, and \code{Actuator}.
We will now explain the code for this program, and, in the process, introduce Lingua Franca.
\autoref{fig:furuta-code} and \autoref{fig:furuta-impl-code} represent an abbreviated version with some C code omitted for clarity.
The source code of a more comprehensive implementation is available online.\footnote{\url{https://github.com/lf-lang/examples-lingua-franca/tree/date23/C/src/modal_models/FurutaPendulum}}

Let us start with the core of the program, the controller itself.
The very first line in \autoref{fig:furuta-code} identifies the target language as C, which means that this controller will be translated into a standalone C module, and that the logic of reactions and mode transitions will be written in C.
The first two lines in the \code{Controller} reactor define the input and output ports.
It has a vector-valued input port named \code{angles} that accepts inputs containing measurements of the angle of the shaft, the angle of the pendulum, and the angular velocities of both, as measured by sensors.
Following are three \emph{reaction} definitions, each reacting to the \code{angles} input, producing a control output, and implementing one of three control laws.

We here use the new mode extension to encapsulate each one in a separate mode.
The reaction bodies are given in ordinary C code that reads the input values and calculates an actuation signal.
That C code would go on lines \ref{ln:cl1}, \ref{ln:cl2}, and \ref{ln:cl3} but is abstracted away here because those details are not germane to this paper.
On lines \ref{ln:out1}, \ref{ln:out2}, and \ref{ln:out3}, the calculated control value is sent to the output port.
Lines \ref{ln:cond1}, \ref{ln:cond2}, and \ref{ln:cond3} use C expressions (abstracted here) to determine whether a mode change is now required,
and, if so, invoke \code{lf\_set\_mode} to specify the next mode.
See \autoref{sec:semantics} for the semantics of these mode transitions.

Next, \autoref{fig:furuta-impl-code} illustrates how this controller is used in the actual program, which is generated into a standalone C program that can be loaded into the flash memory of an embedded controller and deployed.
To interact with the real world, the \code{Sensor} and \code{Actuator} reactors also need to be implemented.
We skip the presentation of these for brevity and only import them as for the \code{Controller}.
The main reactor on line \ref{ln:main}, representing the actual program, instantiates those two reactors along with the imported \code{Controller} and then connects their ports (lines \ref{ln:connect1} and \ref{ln:connect2}).

\begin{figure}
  \centering
  \includegraphics[width=.8\linewidth]{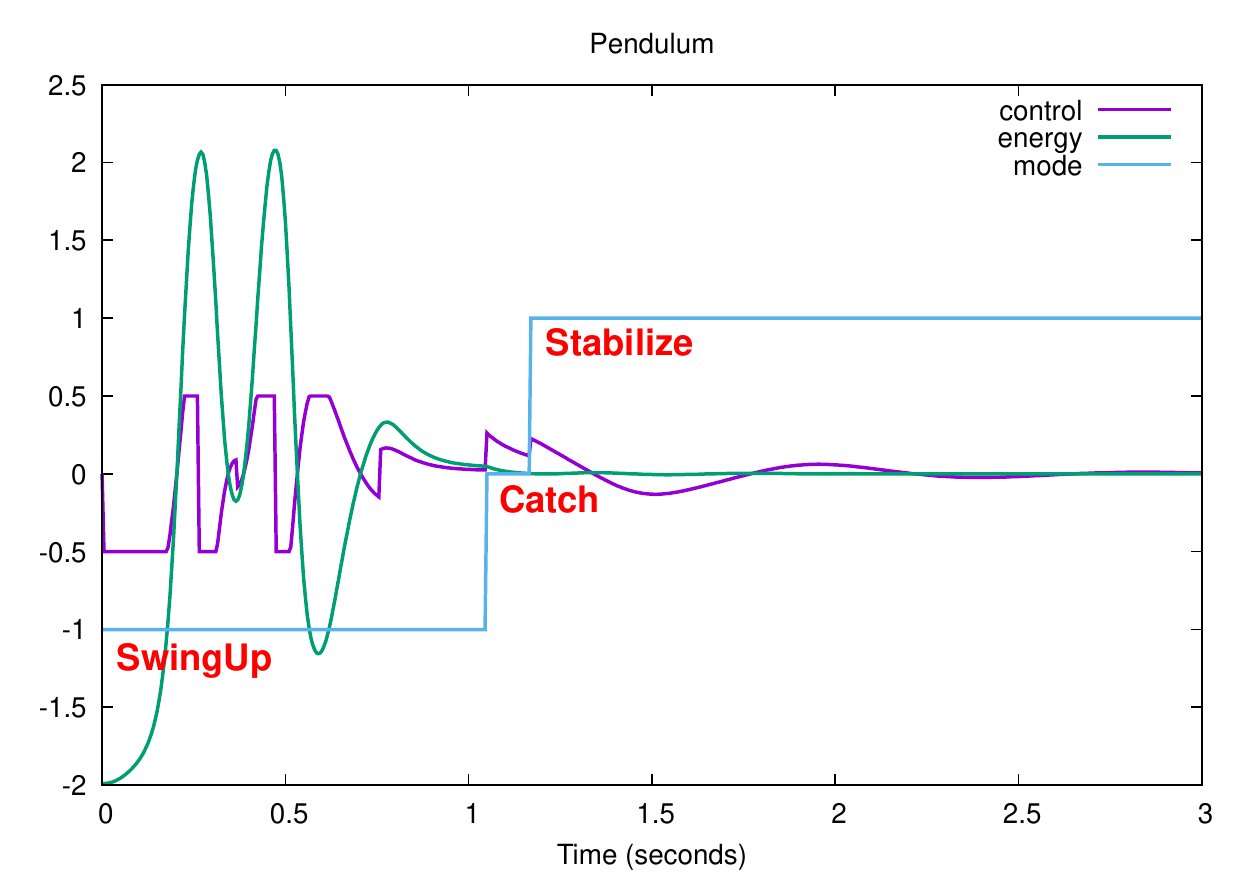}
  \caption{Plot of behavior using \code{Controller} with a simulated pendulum.}
  \label{fig:furuta-plot}
\end{figure}

We used the model given by Eker et al.~\cite{liu2002realistic} to construct a simple forward-Euler simulation of an actual pendulum and ran it with our controller.
The results is plotted in \autoref{fig:furuta-plot}.
This plot shows that the controller spends a little more than one second in the \code{SwingUp} mode, about 100 msec in the \code{Catch} mode, and then remains in the \code{Stabilize} mode.

\begin{figure}
  \centering
  \subfloat[\ac{lf} code]{
    \begin{minipage}[b]{\linewidth}
    \centering
\begin{lstlisting}[language=LF,escapechar=|]
target C;
preamble {=
    typedef enum {SwingUp, Catch, Stabilize} modes;
=}
reactor Controller {
    input angles:double[];
    output control:double;
    state my_mode:modes;
    
    reaction(angles) -> control {=
        if (self->my_mode == SwingUp) {
            ... control law here in C ...
            lf_set(control, ... control value ... );
            if ( ... condition ... ) {
                self->my_mode = Catch;
            }
        } else if (self->my_mode == Catch) {
            ... control law here in C ...
            lf_set(control, ... control value ... );
            if ( ... condition ... ) {
                self->my_mode = Stabilize;
            }
        } else {
            ... control law here in C ...
            lf_set(control, ... control value ... );
            if ( ... condition ... ) {
                self->my_mode = SwingUp;
            }
        }
    =}
}
\end{lstlisting}
    \label{fig:furuta-non-modal-code}
    \end{minipage}
  }

  \subfloat[Diagram]{
    \begin{minipage}[b]{\linewidth}
    \centering
    \includegraphics[scale=\lfscale]{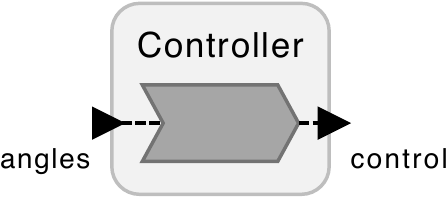}
    \label{fig:furuta-non-modal-diagram}
    \end{minipage}
  }
  \caption{Realization of \code{Controller} without using modal models.}
  \label{fig:furuta-non-modal}
\end{figure}

We admit that this controller is rather simple and could easily be realized without the mode extension we propose in this paper.
Such a realization is sketched in \autoref{fig:furuta-non-modal}.
This variant has a preamble of target-language code to define an enum,
and then, in the body of the reaction to sensor inputs, if-then-else statements are used to determine which control law to invoke.
Although for this example the version without explicit modes is not unappealing, for more complex systems, we believe that the modal version of \autoref{fig:furuta} is more modular and easier to understand.
Moreover, \autoref{fig:furuta-non-modal-code} relies on a hand-written state machine implementation.
While this might be acceptable for just three states, it contradicts the fundamental idea of model-driven engineering and it is also easily prone to errors, complex to extend, and hinders formal verification.
From a modeling perspective the explicit use of modes also results in more meaningful diagrams, comparing \autoref{fig:furuta-non-modal-diagram} to \autoref{fig:furuta-diagram}.
In more advanced modal scenarios, features like mode-local time and history transitions (explained later) add expressiveness that is not easily replicated in non-modal \ac{lf}.

\section{Reactor-oriented Programming}
\label{sec:reactors}

The reactor-oriented programming paradigm that is central to the \ac{lf} coordination language was only introduced recently~\cite{Lohstroh:2019:CyPhy}. 
It is based based largely on principles borrowed from well-established paradigms (such as object-oriended~\cite{OOPStroustrup}, actor-based~\cite{Hewitt:77:Actors}, event-driven~\cite{eventdrivenprog}, flow-based~\cite{Conway:63:Coroutines}, etc.).

To justify why our contribution extends reactors and not some other model of computation, we first discuss the key principles of the reactor model and its advantages that the rest of this paper builds on.
Readers already familiar with that may skip this section.
Conversely, we refer readers interested in more detail, including a formalization, elsewhere~\cite{Lohstroh:EECS-2020-235,LohstrohEtAl:21:Towards}.

\subsection{Composability and Causality}
\emph{Reactors} are in many ways analogous to objects; they are declared in the form of an instantiable class, and they offer a form of inheritance.
Like objects, reactors encapsulate state; all state variables in a reactor are protected, visible to subclasses but otherwise invisible. 
With objects, interaction is primarily via method calls.
In contrast, reactors have reactions, ports, timers, and actions, where a \emph{reaction} is a procedure that is invoked in response to input, timer, or action events.
Unlike a method, a reaction is not invoked directly, but triggered by the presence of an event. 

\emph{Ports} are terminals through which events are relayed between reactors.
Ports of different reactors can be wired together using connections, enabling a flow of events.
\emph{Actions} are very similar to ports, but are private and used only for the scheduling of future events (as opposed to ports, which relay events that are currently present).
There are two kinds of actions: \emph{physical actions}, which are meant to be scheduled from an asynchronous context like an interrupt service routine, and \emph{logical actions}, which should only be scheduled during the course of a reaction. While events scheduled using a logical action is offset with respect to the current tag, events scheduled on a physical action are offset relative to a measurement from a physical clock.
\emph{Timers} are actions that are automatically rescheduled with a predefined periodicity.

Each reaction specifies \emph{triggers}--ports, actions, or timers that can trigger it;
\emph{sources}--ports it can read from when triggered; and \emph{effects}--ports or actions it may produce an event on.
Scoping rules ensure that the data dependencies expressed in the reaction signature are conservative.
Hence, reaction signatures encode a causality interface~\cite{lee2005causality}.

A \emph{composition} of reactors, then, can be turned into a dependencies graph that organizes all reactions into partial order that captures all scheduling constraints that must be observed to ensure that the execution of a reactor program yields deterministic results.
Because this graph is valid irrespective of the contents of the code that executes when reactions are triggered, reactions can be treated as a black box.
It is this property of reactors that enables the polyglot nature of \ac{lf}.

\subsection{Reactivity and Synchronicity}
Control is handed to reactions only when there are events that trigger them.
The runtime mechanism behind can be compared to JavaScript's event loop~\cite{10.1145/3170472.3133846}.
The JavaScript event loop, however, is asynchronous, whereas the execution model of reactors is more like that of the synchronous languages \cite{1173191}, as execution progresses from one synchronous-reactive ``tick'' to the next. 
\emph{Ticks} correspond to monotonically increasing tags, and only events with a matching tag are considered present during a particular tick.
Outputs produced at a given tick are logically instantaneous and may consequently trigger more reactions to occur at the same tick.
Importantly, the runtime system ensures that no reaction executes before all values it may depend on are known.
This is what gives reactors a deterministic semantics.

\subsection{Time and Concurrency}
The \emph{tags} of events delineate a logical timeline, but are also used as a means to keep the system synchronized with physical time.
Tags are pairs $(t,m)$, where $t$ is a time value and $m$ a microstep index, the latter of which is only used to enable subsequent ticks to occur without any time elapsing between them, a concept known as \emph{superdense time}~\cite{manna1992verifying}. 
The execution engine starts a new tick once the time value of the earliest available event tag is greater than the current physical time. 
This policy ensures that events scheduled via a \emph{physical action} (from an asynchronous thread of execution) can be tagged with the current physical time without risking out-of-order event tags. 
The relationship between physical and logical time in the reactor model gives timers a useful semantics and also permits the formulation of deadlines~\cite{lohstroh2020language}. 
These constructs are particularly practical for software that operates in cyber-physical systems (like the Furuta pendulum).

Because the execution of reactors is guided by a dependency graph, logically simultaneous reactions without dependencies between them can transparently be executed in parallel without introducing any data races, deadlocks, or other common problems that crop up when multiple threads of execution are involved~\cite{Lee:06:Threads}.
The current runtime implementations in C, C++, and Rust automatically exploit instances of such parallelism (if the execution platform has multiple cores).
It should be noted that reactions within the same reactor always have dependencies between them to ensure mutually exclusive access to shared state. Specifically, the order of declaration in the code determines precedence.

\subsection{Visualization and Pragmatics}
\label{sec:pragmatics}
The reactor-oriented programming paradigm with components, ports, connections, and hierarchy, lends itself particularly well to visualizations that can aid programmers in understanding code. Programming models that are not flow-based are amenable to control-flow analysis that can deliver graphical renditions to help explain program behavior, but it is often difficult to determine a suitable scope or grain for such analysis, making them difficult to automate. More common are visualization that provide insight in a high-level software architecture (e.g., class diagrams) but, while easy to automate, these reveal little about actual program behavior. 

\ac{lf} comes with an automatic interactive diagram synthesis capability, tailored to enhance developer's grasp of their code and increase their productivity.
One could say that \ac{lf} capitalizes on ``pragmatics''~\cite{vonHanxledenLF+22} when it comes to the handling of models of possibly highly complex systems, with a focus on how to get the best of both the textual and the graphical worlds. The extension presented in this paper leverages this capability and augments it with visualizations tailored to modal reactors. All the \ac{lf} diagrams, including the ones presented in this paper, are synthesized automatically from \ac{lf} code using a port-aware variant of the well-known Sugiyama algorithm~\cite{SchulzeSvH14}, provided by the Eclipse Layout Kernel (ELK)\footnote{\url{https://www.eclipse.org/elk/}}.

\section{Modal Reactors}
\label{sec:modal-reactors}

The basic idea of modal reactors is to use the existing reactor model but to allow for a modal coordination of reactions, by partitioning reactors into disjoint subsets that are associated with mutually exclusive \emph{modes}.
In a modal reactor, only a single mode can be active at a particular logical time instant, meaning that activity in other modes is automatically suspended.
\emph{Transitioning} between modes switches the reactor's behavior and controls the starting point of the entered modes.
The option to reset or continue with the mode's history are common and powerful abstractions that are particularly helpful in managing complex timed behaviors, which can be extremely error-prone when carried out manually.

While the ideas behind modal reactors are not new, the guidance by \ac{lf}'s fundamental principles towards a polyglot modal coordination layer is novel.
Our goal is to create modal models that are:
\begin{itemize}
   \item \textbf{lean}\enspace a minimal coordination layer that provides the most essential functionality but still offers maximal versatility and user adjustability;
   \item \textbf{polyglot}\enspace a flexible multi-language wrapper that focuses on the user's language and requires only minor adaptation effort;
   \item \textbf{concurrent}\enspace allowing the design of multiple separate modal units acting independently;
   \item \textbf{timed}\enspace a reliable and precise way to specify time sensitive modal behavior, even in parallel and distributed environments; and
   \item \textbf{deterministic}\enspace yielding unambiguous and reproducible output behavior for the same sequence of tagged input events.
\end{itemize}
Our modal extension to \ac{lf} embodies these very principles and embraces the crucial ``black box'' approach to reactions.

\subsection{Syntax}
\label{sec:basic-syntax}

The additional syntax required for adding modes is rather minimal.
The added syntax allows reactions to be grouped into modes and includes new keywords for declaring (initial) modes and specifying transition types.
The core \ac{lf} language remains unchanged.

Modes can be defined in any reactor.
Each mode requires a unique (per reactor) name and can declare contents that are local to this mode.
There must be exactly one mode marked as \code{initial} (see line~\ref{ln:swingup} in \autoref{fig:furuta}).
A mode can contain state variables, timers, actions, reactions, reactor instantiations, and connections.
While the modes cannot be nested in other modes directly, hierarchical composition is possible through the instantiation of modal reactors.
The main exception in allowed contents in modes are port declarations, as these are only possible on reactor level.
Yet, modes share the scope with their reactor and, hence, can access ports, state variables, and parameters of the reactor.
Only the contents of other modes are excluded.

Mode transitions are declared within reactions.
If a reactor has modes, reactions are allowed to list them as effects.
This enables the use of the target language API to set the next mode, using \code{lf\_set\_mode} (see also line~\ref{ln:mt1} in \autoref{fig:furuta-code}).
The compiler will reject the program if the target code references a mode that is not declared as an effect.
The user also has to specify the type of the transition by adding the modifier \code{reset} or \code{history} to the effect. 
An effect declared as \code{history(<mode>)} specifies a \emph{history transition} to the mode, rendered in the graphical syntax with an ``H'' at the arrowhead (see \autoref{fig:local-time-model}).

\subsection{Modes and Transitions}
\label{sec:semantics}

The basic effect of modes in \ac{lf} is that only parts that are contained in the currently active mode, or not contained in any mode, are executed at any point in time.
This also holds for parts that are nested in multiple \emph{ancestor modes} due to hierarchy; consequently, all those ancestors must be active in order to execute.
Reactions in inactive modes are simply not executed.
All components that model timing behavior, namely timers, scheduled actions, and delayed connections, are subject to a concept of \emph{local time}.
That means while a mode is inactive, the progress of time is suspended locally.
How the timing components behave when a mode becomes active depends on the transition type.
A mode can be \emph{reset} upon entry, returning it to its initial state.
Alternatively, if it was active before, it may continue based on its \emph{history}.
In the latter case all timing components will continue their delays or period as if no time had passed during inactivity of the mode.
\autoref{sec:local-time} will provide further insights to the concept of local time.

Upon reactor startup, the initial mode of each modal reactor is active, others are inactive.
If at a tag $(t, m)$, all reactions of this reactor and all its contents have finished executing, and a new mode was set in a reaction, the current mode will be deactivated and the new one will be activated for future execution.
This means no reaction of the newly active mode will execute at tag $(t, m)$;
the earliest possible reaction in the new mode occurs one microstep later, at $(t, m+1)$.
Because of its superdense time model, \ac{lf} is able to model a subsequent reaction at the same logical time, but one microstep later.
Hence, if the newly active mode has for example a timer that will elapse with an offset of zero, it will trigger at $(t, m+1)$.
In case the mode itself does not require an immediate execution in the next microstep, it depends on future events, just as in the normal behavior of \ac{lf}.
Hence, modes in the same reactor are always mutually exclusive \wrt superdense time.
\autoref{sec:timing-transitions} will discuss more details about this design.

A transition is triggered if a new mode is set in a reaction body, as done on lines \ref{ln:mt1}, \ref{ln:mt2}, and \ref{ln:mt3} of \autoref{fig:furuta-code}.
As with setting output ports in reaction, a new mode can be set multiple times in the same or different reaction.
In the end, the fixed ordering of reactions determines the last effective value that will be used.
The new mode does not have to be a different one; it is possible for a mode to reset itself via a reset transition.

In case a mode is entered with the reset behavior,
  all contained modal reactors are reset to their initial mode (recursively),
  all local timers are reset and start again awaiting their initial offset,
  all events (actions, timers, delayed connections) that were previously scheduled from within this mode are discarded, and
  a newly introduced \code{reset} trigger activates associated reactions in the mode and all contained reactors (recursively).

Thus, whenever a mode is entered with a reset transition, the subsequent timing behavior is as if the mode was never executed before.
By default, state variables are not reset automatically because it is idiomatic for reactors to allocate resources or initialize subsystems 
(e.g., allocate memory or sockets, register an interrupt, or start a server) 
in reactions triggered by the \code{startup}, and to store references to these resources in state variables.
Consequently, if \code{startup}, as well as \code{shutdown}, reactions
are used inside modes, they are subject to special handling when it comes to
mode activity, to ensure correct allocation and deallocation of resources.
\autoref{sec:startup} will elaborate on this topic.
In any case, if there are state variables that need to be reset or reinitialized, then this can be done in a reaction triggered by \code{reset} or by marking their declaration with \code{reset} to automate the assignment of an initial value.

On the other hand, if a mode has been active prior and is then re-entered via a history transition, no reset is performed.
Events originating from timers, scheduled actions, and delayed connections are adjusted to reflect a remaining delay equal to the remaining delay recorded at the instant the mode was previously deactivated.
As a consequence, a mode has a notion of local time that elapses only when the mode is active, see \autoref{sec:local-time}.

\subsection{Timing of Transitions}
\label{sec:timing-transitions}
An important design decision for modes in \ac{lf} regards the timing of transitions, i.e., when exactly a transition takes place.
There are different established semantics, the two basic alternatives being an ``immediate'' transition or a ``delayed'' transition where a target mode gets enabled at least one microstep after the transition is initiated.

With an immediate effect, transitions could occur directly after executing the initiating reaction, which would instantly activate reactions in the target mode.
This implies dependencies between reactions with transition effects and reactions in target modes.
Such additional dependencies could lead to causality loops in otherwise legitimate modal models and could reduce exploitable parallelism.
An alternative for timing of transitions could be to wait until all contents of a mode finished executing but then immediately switch to the next mode and execute that one still at the same tag.
However, this would raise the question of how to handle the reactivation of the same mode multiple times at the same tag.
Moreover, this would allow an arbitrary number of mode changes during the same execution instant.
In the end, some notion of a sequential order between mode activations would be necessary, which is essentially what microsteps are used for in our delayed variant.

In \ac{lf}, reactions can set a new mode but this has no immediate effect.
Only when the reactor finished executing all its contents, the transition is performed and the new mode becomes active in the next microstep.
With this design, no two modes in the same reactor can be active at the same tag.
Neither can a transition interfere with ongoing reactions.
Yet, this approach requires resolving potentially ``conflicting'' transition effects from different reactions.
Here, we apply the same mechanism as already used in setting ports in \ac{lf} and determine the effective target mode by relying on the fixed ordering of reactions within a reactor.
In terms of deterministic outcome and overriding behavior, setting new modes can be considered analogous to assigning output ports.
However, in terms of timing, transition effects correspond to scheduling actions with a zero delay, which also enforce a microstep delay to prevent causality cycles.

The design we chose is similar to the design choice in \ac{scade}~\cite{Colaco:17:SCADE}, where states within a reaction are mutually exclusive.
It differs from the design choice in SyncCharts~\cite{Andre:04:SyncCharts} or SCCharts~\cite{vonHanxledenPLDI14}, where multiple states may be activated within one reaction by a sequence of ``immediate transitions.''
Note that both choices achieve a deterministic semantics. In the case of SCCharts, determinism is ensured by means of a predetermined sequential ordering of states.

\subsection{Local Time}
\label{sec:timing-discussion}\label{sec:local-time}

The notion of mode-local time, with the suspension of all timing behavior within inactive modes, is an established and well-formed principle also found in modal models in Ptolemy II~\cite{LeeTripakis:10:ModalModels}.
The considerations there that favor local time over alternative approaches also apply to \ac{lf}.
The suspension of time gives a clear and consistent meaning to the inactivity of modes and provides a comprehensible state for the mode's contents upon entry.
This especially favors modularity, as reactors that may be instantiated in modes do not have to anticipate the fact that their time (driven by timers or scheduled actions) will advance while their reactions are suppressed.
Furthermore, modes allow to define reactor elements outside of modes, which gives the developer control over whether time should be local to a mode or not.

\begin{figure*}
  \centering

  \hfill
  \subfloat[The \ac{lf} Model]{
    \includegraphics[scale=\lfscale]{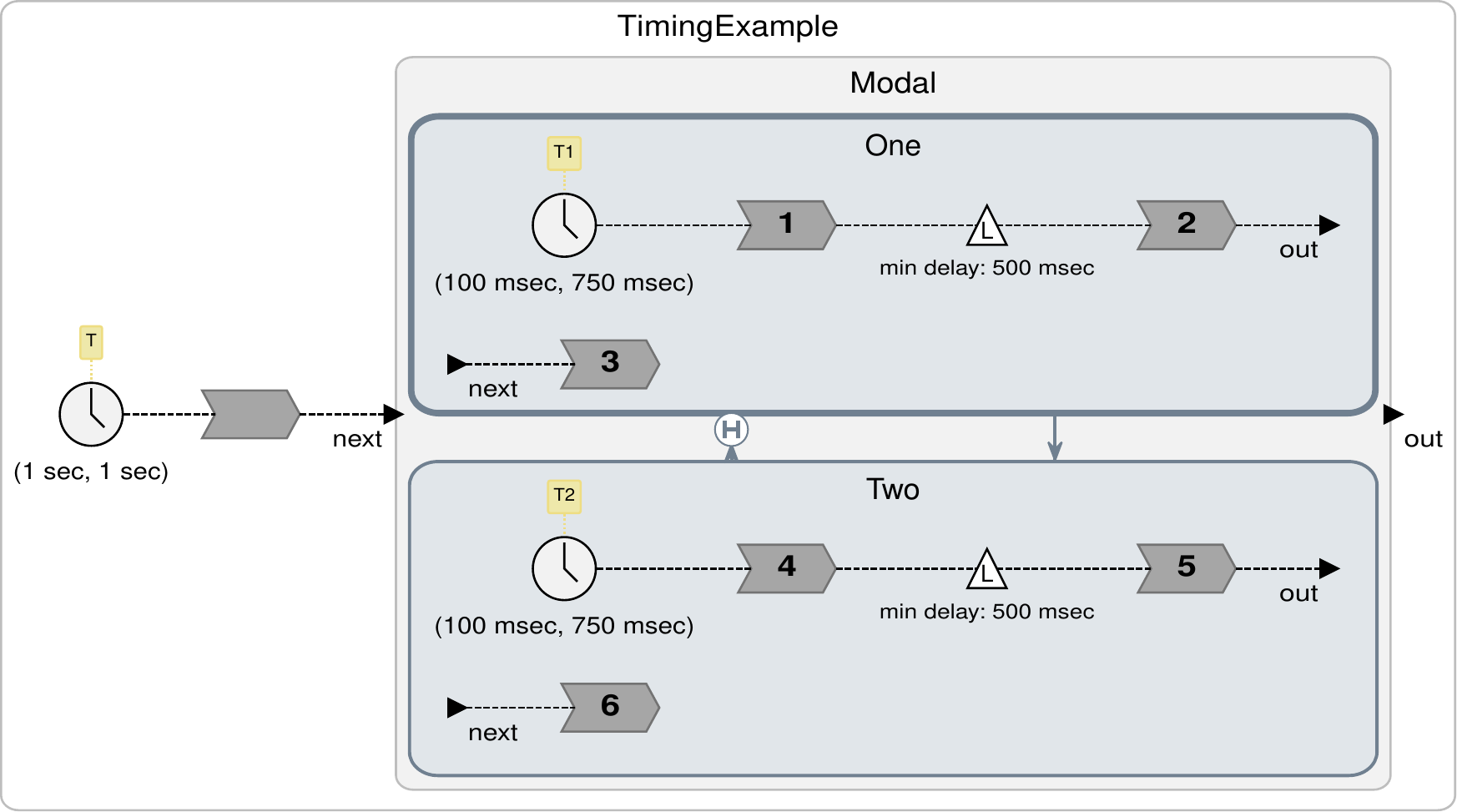}
    \label{fig:local-time-model}
  }
  \hfill
  \subfloat[The progression of time in each mode and their respective timer]{
    \hspace{7mm}
    \includegraphics[scale=.58]{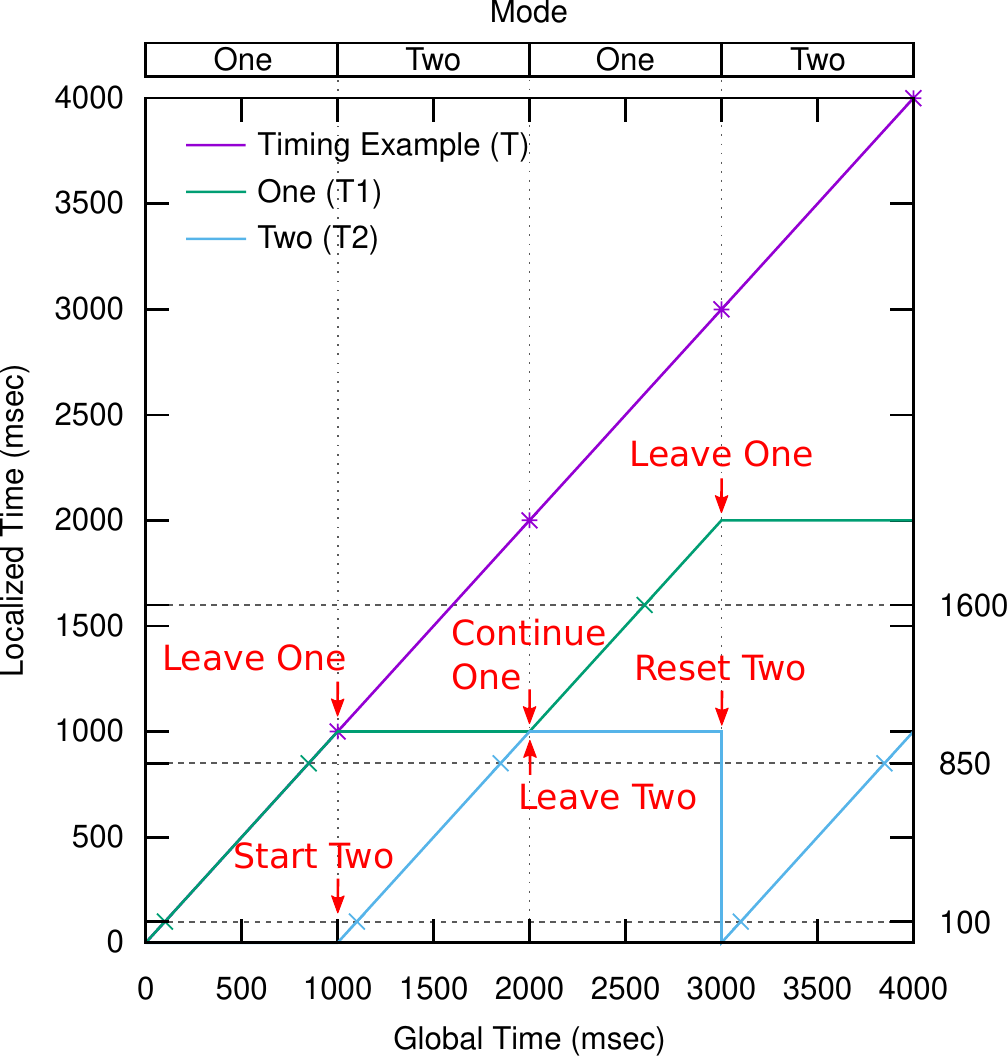}
    \hspace*{7mm}
    \label{fig:local-time-plot}
  }
  \hfill\\
  \subfloat[The execution trace with reaction illustration]{
    \includegraphics[width=.99\linewidth]{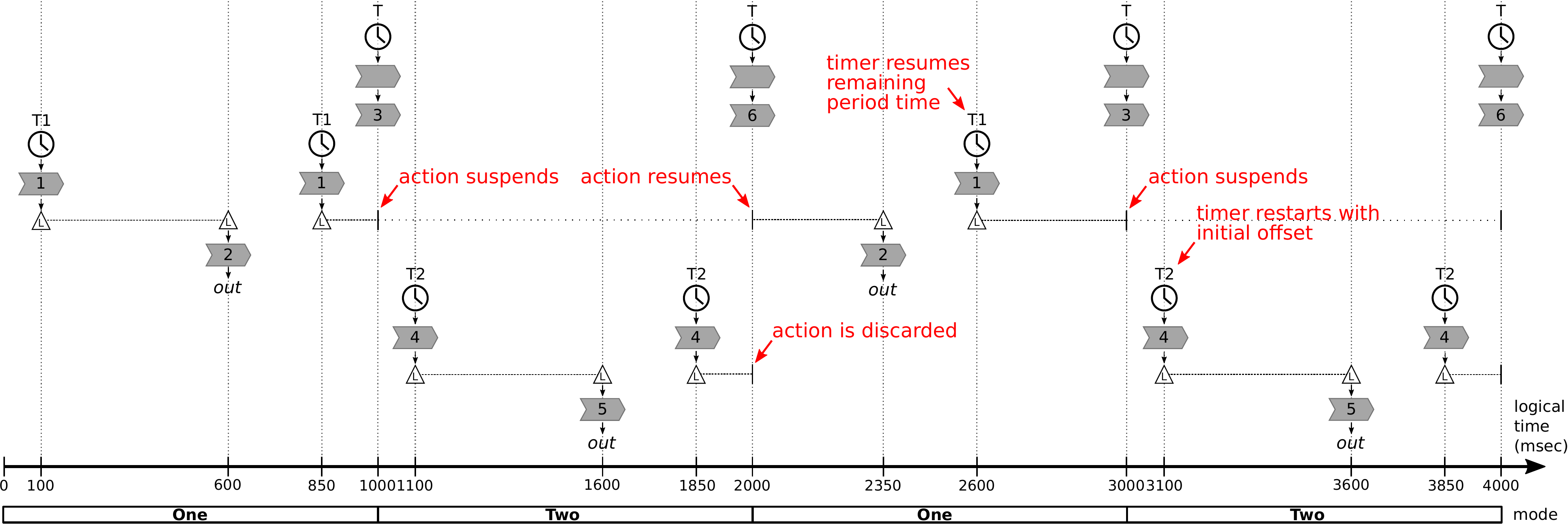}
    \label{fig:local-time-trace}
  }

  \caption{\ac{lf} example illustrating the different effects of reset and history transitions on timers and delays in modes.}
  \label{fig:local-time}
\end{figure*}

\autoref{fig:local-time} illustrates the different characteristics of local time affecting timers and actions in the presence of the two transition types.
\autoref{fig:local-time-model} shows the generated diagram for a synthetic example program.
It consists of two modes \textsf{One} (the initial mode) and \textsf{Two}, both in the \textsf{Modal} reactor.
The \textsf{next} input toggles between these modes.
The input port is controlled by a reaction at the top level that is triggered by the timer \textsf{T}.
After one second, a mode switch is triggered periodically with a period one second.
The modes' contents are structured identically.
Each has a timer \textsf{T1}/\textsf{T2} that triggers a reaction after an initial offset of 100 msec and then periodically after 750 msec.
This reaction then schedules a logical action with a delay of 500 msec (the actual target code does not add an additional delay over the minimum specified).
This action triggers the second reaction, which writes to the output \textsf{out}.
The last reaction is triggered by the input \textsf{next} and invokes the transition to the other state.
The main difference between the modes is that \textsf{One} is entered via a history transition, continuing its behavior, while \textsf{Two} is reset.

\autoref{fig:local-time-trace} illustrates the execution trace of the first 4 seconds of this program.
Below the timeline is the currently active mode and above the timeline are the
model elements that are executed at certain points in time, together with arrows
indicating triggering relations and dashed lines for distribution through time.
For example, at 100 msec, the initial offset of timer \textsf{T1} elapses, which leads to the scheduling of the logical action in this mode.
The action triggers the reaction 500 msec later, at 600 msec, and thus causes an output.
The timing diagram illustrates the different handling of time between history transitions and reset transitions.
Specifically, when mode \textsf{One} is re-entered via a history transition, at time 2000 msec, the action triggered by \textsf{T1} before, at time 850 msec, resumes.
In contrast, when mode \textsf{Two} is re-entered via a reset transition, at time 3000 msec, the action triggered by \textsf{T2} before, at time 1850 msec, gets discarded.

\autoref{fig:local-time-plot} illustrates the relation between global time in the environment and the localized time for each timer in \autoref{fig:local-time}.
Since the top-level reactor \textsf{TimingExample} is not enclosed by any mode, its time always corresponds to the global time.
Mode \textsf{One} is the initial mode and hence progresses in sync with \textsf{TimingExample} for the first second.
During inactivity of mode \textsf{One} the timer is suspended and does not advance in time.
At 2000 msec it continues relative to this time.
\textsf{T2} only starts advancing when the mode becomes active at 1000 msec.
The reentry via reset at 3000 msec causes the local time to be reset to zero.

\medskip
In short, from the perspective of timers and actions, time is suspended when a mode is inactive.
This also applies to indirectly nested reactors within modes.
In the same way, connections with logical delays (given in \ac{lf} by the \code{after} keyword) are affected by local time, if their source lies within a mode.
This corresponds to that fact that delayed connections are to a certain degree syntactic sugar for connections delayed by a logical action.

\subsection{Startup and Shutdown}
\label{sec:startup}

One design challenge we faced was figuring out how to handle start and end of program execution.
\ac{lf} defines two special triggers, \code{startup} and \code{shutdown}.
A reaction triggered by \code{startup} will execute during the very first tag at which the reactor exists, and one triggered by \code{shutdown} will execute during the very last tag before the reactor ceases to exist.
These are commonly used to set up a reactor, for example to allocate memory for state variables,
initial sensors or actuators, or start external threads (e.g., to listen for network inputs).
The \code{shutdown} reactions subsequently clean up before exiting.
What if such reactions are found within a mode?
Should these be executed upon entering or exiting a mode?

We considered several different options and came to the conclusion that all of them had some oddities and strange corner cases.
We have implemented a strategy that is simple to explain, deterministic, and enables workarounds for all the oddities we have been able to identify,
but we admit a level of dissatisfaction.
Our solution is a compromise.

First, we invoke \code{startup} reactions at most once at the first activation tag of a mode.
Second, we invoke \code{shutdown} reactions at the last tag before the containing reactor ceases to exist (usually just prior to the program exiting),
irrespective of mode, but we only invoke those reactions in modes that have activated at least once.
Hence, every \code{startup} has a corresponding \code{shutdown}.
Third, as explained earlier, we introduced a new \code{reset} trigger for reactions.
Reactions triggered by \code{reset} will execute each time a mode is entered via a reset transition.

Consider, for example, memory allocation and deallocation.
These can now be done safely within a mode in \code{startup} and \code{shutdown} reactions, although this strategy will have the disadvantage
that memory does not get deallocated while a mode is inactive.
To allocate memory for use only while a mode is active, a programmer could allocate the memory in a \code{reset} reaction
and deallocate it in any reaction that calls \code{lf\_set\_mode} to exit the mode.
The programmer would then have to ensure that every entering transition is a reset transition.

An oddity that results from our design is that \code{shutdown} reactions are invoked even in inactive modes.
Normally, no reaction of an inactive mode should be invoked.
This is the only exception, but it ensures that every triggering of a \code{startup} reaction is matched by a corresponding triggering
of a \code{shutdown} reaction.

An alternative solution, executing shutdown upon leaving a mode, makes history transitions infeasible.
We also considered scattering \code{shutdown} reactions across multiple consecutive microsteps so that no two distinct modes had logically simultaneous reactions.
This seemed difficult to explain and likely to lead to even more peculiar oddities.
A third possibility we considered was executing all startup reactions at the beginning, also disregarding whether a mode is active.
But this makes \code{startup} reactions as strange as \code{shutdown} reactions, and we felt it was better to have fewer strange behaviors.

\subsection{Mode Diagrams}
\label{sec:diagrams}

As mentioned in \autoref{sec:pragmatics}, we consider (automatic) diagramming to be an important aspect of \ac{lf} and also adopt modal behavior as a first-class citizen to this graphical notation.
Even if the actual triggering of a mode change is in the verbatim target code and not part of the \ac{lf} language itself, the declaration of modes and potential transitions in the reaction interface is sufficient to provide the user with mode diagrams.
However, one may specify state transitions at the \ac{lf} level that do not actually exist in the host code (but not vice versa as explained in \autoref{sec:basic-syntax}).
Thus, mode diagrams are conservative in that they show all possible behaviors, but they may show transitions that are not realized in the host code.

An important design decision for \ac{lf} diagrams is how data flow and control flow should visually relate to each other.
In \ac{lf}, reactors are denoted with rounded rectangles and their data flow is visualized with rectangular edge routing.
Following established practices for state machine models, modes are added as rounded rectangles as well but with a different color scheme.
Initial states are indicated with a thicker outline.
\autoref{fig:furuta-diagram} illustrates this notation.
State transitions are drawn as splines.
For transitions that continue the behavior in the mode as in \autoref{fig:local-time-model}, we add a circled ``H'' in front of the arrow to represent that this mode will be entered including its previous history.

In contrast to other \ac{lf} diagrams in this paper, the transitions in \autoref{fig:furuta-diagram} also feature labels.
\ac{lf} diagrams offer various ways for a user to influence the appearance and level of detail in diagram.
Accordingly, modes also have some configuration options, such as the presence of labels.
The labels can provide additional value because they indicate which inputs and/or actions could cause a mode transition to be taken.
Again, labeling is subject to certain design considerations.
Traditionally, transition labels include triggers and effects.
However, this would require an analysis of host code contradicting the polyglot approach, and the conditions that actually lead to taking a transition, as well as the effects that result from taking that transition, might become arbitrarily complex.
We therefore opted to restrict the transition labels to the events that \emph{may} trigger a reaction, omitting whatever further logic inside the host code determines whether a transition will actually be taken.
If labels are configured to not be present and leading to an absence of any distinguishing factors, we bundle multiple transitions between the same modes into one.

Another, less obvious question is how to integrate reactions dependencies into the new layer of modes and transitions.
One option would be to try to mix all these edges.
However, this would mean that data flow edge may need to cross the hierarchy level of the mode to connect any content of the mode.
We created several visual mockups for that.
All variants that included some form of cross-hierarchy edges, were considered quite messy as soon as they exceeded a trivial size.
Additionally, the interaction for collapsing modes to hide their contents and the feasibility with respect to automatic layout algorithms was considered.
In the end, we opted for breaking up these outside connections on the level of each mode.
This makes some connections more implicit but leads to cleaner diagrams and simplifies the layout problem.
For ports, we duplicate those used in a mode and represent them by their arrow figure.
The name is used to create an association to the original reactor port.
In \autoref{fig:furuta-diagram} you can see this in the \textsf{angles} input triggering reactions.
For connections to reactor elements that are defined outside of modes, \eg a timer,
we create artificial port-like elements to illustrate their relation.

\section{Implementation}
\label{sec:implementation}

The goal of creating a minimally invasive mode extension is not only reflected in the language design itself but also in its runtime implementation.
\ac{lf} with its polyglot approach supports various target languages.
At the time of writing this paper, modes are implemented for C and Python, but extending this is already planned for future development.
Nonetheless, modes have already been successfully tested for these targets with single-threaded, multi-threaded, and distributed execution.
For the remaining course of this section, we will provide a generic and reasonably target language-independent view on the adjustment required to support modes, achieving the behavior described in \autoref{sec:modal-reactors}.
The actual implementation can be found in the \ac{lf} project on GitHub.\footnote{\url{https://github.com/lf-lang}}

Code generated by \ac{lf} consists of two parts.
First, the program-specific parts that represent the components and topology of the program itself, such as reactors, their actual instantiation, and interconnection.
The second part is a generic runtime environment that among other things handles processing and scheduling of events in the event queue or triggering and execution of reactions.
The adjustments for extending both parts towards modes are relatively small.
Moreover, they are mostly additive, which means that in the absence of modes in the source model, there is virtually no difference to an implementation that does not account for modes\footnote{Some data structures require references to a potentially enclosing mode. This is necessary to enable the use of pre-compiled reactors in modal reactors. However, this only results in a minimal memory overhead of pointers that remain unused in the absence of modes.}.

The existing runtime implementation needs to be adjusted in two ways.
First, the triggering of reactions must check if a reaction is in an active mode and otherwise prevent its execution.
A trivial approach is to recursively check a mode and all its parent modes whether they are all the currently active mode, while any element associated with no mode at all is considered always active.
Shutdown reactions are excluded from this activity check but their mode (if any) must have had a startup phase, which is additionally recorded when executing these reactions.
Second, the execution life-cycle requires the handling of transitions.
It must be invoked after processing of reactions has finished but before the logical time advances.

\autoref{alg:transition} presents the procedure that handles mode transitions, which includes performing resets, managing local time, and scheduling special triggers.
The algorithm relies on a few global data structures:
(1) the existing event queue of \ac{lf} (EventQueue), which manipulation is the sole change to the runtime that is needed to implement local time;
(2) a collection to store events suspended in local time (SuspendedEvents); and (3) a set of all modal reactor instances in the model (ModalReactors).
The latter is a result of program-specific generated code, as well as a few new data structures and references.
Each modal reactor $r$ provides access to the following information (presented in a member notation here).
$r_{modes}$ denotes the set of modes in $r$.
\begin{algorithmic}
\State $r$.parentMode $\in \{$Nil$\} \cup \{ x_{modes} : x \in \text{ModalReactors} \}$
\State $r$.initialMode $\in r_{modes}$
\State $r$.currentMode $\in r_{modes}$
\State $r$.nextMode $\in \{$Nil$\} \cup r_{modes}$
\State $r$.transition $\in \{$None, Reset, History$\}$
\end{algorithmic}
$r$.parentMode is either absent or the mode immediately containing $r$.
Note that this models only a unidirectional relation for mode hierarchy.
While one could also consider introducing a list of contained modal reactors, we kept this notation to stick close to our actual implementation, despite structural consequences for the algorithm.
$r$.initialMode is the mandatory initial mode.
As the parent mode, both values are constant and set up at program start.
$r$.currentMode is the currently active mode \wrt to $r$, starting with the initial mode.
Whether the current mode is actually active \wrt execution depends additionally on parent modes.
$r$.nextMode and $r$.transition represent the presence, type, and target of a transition.
These fields are filled if the target code sets a new mode, \eg in line \ref{ln:mt1} of \autoref{fig:furuta-code}.
The transition type is actually inferred from the effect definition.
Furthermore, each mode $m$ in $r_{modes}$ carries additional mode-specific information.
\begin{algorithmic}
  \State $m$.reactor $\in \text{ModalReactors}$
  \State $m$.leaveTime $\in \mathbb{T}$
  \State $m$.reset $\in \{ \text{True}, \text{False} \}$
  \State $m$.hadStartup $\in \{ \text{True}, \text{False} \}$
\end{algorithmic}
$m$.reactor is a constant reference to the mode's reactor.
$m$.leaveTime stores the logical time at which this mode was last left.
It is initialized with the start time of the execution.
$m$.reset indicates that this mode needs to be reset as soon as it becomes active (initially False). 
$m$.hadStartup is a boolean flag that is set from False to True as soon as the mode is active for the first time.
Finally, each reaction, timer, and action also has a reference to its immediately enclosing mode, if any exist.
This allows associating events with modes via their trigger.

\begin{algorithm}
\caption{Processing of mode transitions at the end of each execution cycle.}
\label{alg:transition}
\algrenewcommand\algorithmicindent{.8em}%
\begin{algorithmic}[1]
\ForEach{$r \in $ topdown(ModalReactors)} \label{alg:transition:loop1}

  \If {$r$.parentMode $\neq$ Nil \textbf{and}\\\hspace{3.2em} $r$.parentMode.reactor.transition $=$ Reset}\label{alg:transition:hierarchy-reset-check}
      \State $r$.nextMode := $r$.initalMode
      \State $r$.transition := Reset
  \EndIf
\EndFor

\ForEach{$r \in$ ModalReactors}
  \label{line:alg:transition:loop2}

  \If {$r$.transition $\neq$ None}

    \ForEach{$e \in$ SuspendedEvents}
      \If {$e$.mode $=$ $r$.nextMode}
        \State Remove $e$ from SuspendedEvents
        \If {$r$.transition $=$ Reset}
          \If {$e$ \textbf{is} Timer}
            \State $t$ := shift(currentLTime, $e$.timer.offset)\label{alg:transition:timer-restart}
            \State Insert $e$ into EventQueue with tag $t$
          \EndIf
        \Else \label{alg:transition:resume-events}
          \State $t$ := shift(currentLTime, \label{alg:transition:event-continue}\\\hspace{4.5em} $e$.tag - $e$.mode.leaveTime)
          \State Insert $e$ into EventQueue with tag $t$
        \EndIf
      \EndIf
    \EndFor

    \If {$r$.transition $=$ Reset}\label{alg:transition:update}
      \State $r$.nextMode.reset := True \label{alg:transition:reset-mark}
    \EndIf

    \State $r$.currentMode.leaveTime := currentLTime \label{alg:transition:leave-time}
    \State $r$.currentMode := $r$.nextMode
    \State $r$.nextMode := Nil
    \State $r$.transition := None \label{alg:transition:clear}
  \EndIf \label{alg:transition:update-end}

  \If {isActive($e$.currentMode)} \label{alg:transition:special}
    \If {\textbf{not} $r$.currentMode.hadStartup}
      \State Trigger \code{startup} reactions in $r$.currentMode\\\hspace{3.2em} at the next microstep \label{alg:transition:trigger-startup}
    \EndIf
    \If {$r$.currentMode.reset}
      \State $r$.currentMode.reset := False
      \State Trigger \code{reset} reactions in $r$.currentMode\\\hspace{3.2em} at the next microstep \label{alg:transition:trigger-reset}
      \State Reset state variables in $r$.currentMode that are\\\hspace{3.2em} marked with \code{reset} \label{alg:transition:state-reset}
    \EndIf
  \EndIf
\EndFor

\ForEach{$e \in$ EventQueue} \label{alg:transition:suspend-events-start}
  \If {\textbf{not} isActive($e$.mode)}
    \State Remove $e$ from EventQueue
    \State Add $e$ to SuspendedEvents
  \EndIf
\EndFor \label{alg:transition:suspend-events-end}
\end{algorithmic}
\end{algorithm}

In the first line of \autoref{alg:transition}, every modal reactor instance is processed in a top-down order.
This refers to the partial order of mode hierarchy and ensures that if a mode is entered with a reset, inner modal reactors (line \ref{alg:transition:hierarchy-reset-check}) transition to their initial state with a reset transition recursively.
Afterwards, transitions are processed in a separate iteration.
This iteration is separated from the previous iteration because the hierarchical reset relies on the presence of transition information in parent modes to reset itself accordingly (line~\ref{alg:transition:hierarchy-reset-check}) and this information is now overwritten (line~\ref{alg:transition:clear}).
First, events of the next mode that are suspended in time are processed.
At a reset, all timers are restarted with the initial offset relative to the current time (line~\ref{alg:transition:timer-restart}).
Other events (\eg scheduled actions) are dropped.
For reintroducing previously suspended events into the event queue, the shift function is used to create the correct tag \wrt superdense time:
$\text{shift}(\text{base} : (t,m), \text{offset} : (t,m))$
$$
\begin{array}{c}
=
\begin{cases}
    \text{offset}_t > 0: & (\text{base}_t + \text{offset}_t, \text{offset}_m)\\
    \text{offset}_t = 0: & (\text{base}_t, \text{base}_m + \text{offset}_m + 1)
\end{cases}
\end{array}
$$
This creates a tag that is the base tag shifted by a given offset into the future.
It takes into account that a zero delay offset (\wrt the timestamp $t$) results in a future (incremented) microstep.
In case time should continue due to a history transition (starting line \ref{alg:transition:resume-events}), all events are reintroduced into the event queue with an adjusted target time.
Here, the time that mode was left is subtracted from the original tag (time to happen) of the event, to get the remaining time at time of leaving, which is used to offset this event from the current time (line~\ref{alg:transition:event-continue}).
Next, the actual effect of the transitions is applied to the internal data structures (lines~\ref{alg:transition:update} to~\ref{alg:transition:update-end}).
This includes marking the mode for reset if necessary,
storing the time the mode was left,
setting the new mode, and clearing transition information for use in future execution.
Afterwards, the special reactions are triggered for the current mode.
Note that this takes effect based on mode activity and not triggered by a transition (line~\ref{alg:transition:special}).
The isActive function relies on the definition presented before.
If the mode was never active before, its startup reaction will be triggered at the next microstep.
This includes reactions in the mode and in all inner non-modal reactors.
Likewise, reset reactions are triggered when the mode is marked for a reset.
Additionally, the automatic reset for the respective state variables is invoked, and the flag is cleared.
Finally, all events that are associated with now inactive modes are pulled from the event queue and stored in the suspended events collection (lines \ref{alg:transition:suspend-events-start}--\ref{alg:transition:suspend-events-end}).

In the real implementation of \autoref{alg:transition}, the procedure includes some additional consistency checks, optimizations, and handles for some minor corner cases.
However, as these are not relevant for the overall semantics, we left them out here for ease of readability. 
Overall, the algorithm is fairly compact while providing the bulk of semantic functionality for modes.

\section{Discussion}
\label{sec:discussion}

\subsection{Strong vs.\ Weak Preemption}
\label{sec:preemption}

Mode transitions typically imply a \emph{preemption} of some behavior associated with the mode that is left.
Synchronous statechart dialects, such as SyncCharts and SCCharts, distinguish between a \emph{strong preemption}, where such inner behavior is not executed anymore in the tick that the preemption takes place, and \emph{weak preemption}, where modes still execute a ``last wish'' in that tick.
Thus, it would for example in SyncCharts constitute a compile-time causality error if a state could emit a signal that would strongly preempt that state.

Technically, transitions in modal reactors are weak, in the sense that even when a mode change takes place, all reactions still get to execute at the current tick.
One argument for the absence of strong preemption in the synchronous sense at the \ac{lf} level is that it is the modes themselves that determine transitions to other modes, leading to the aforementioned causality issues in case of a strong preemption.
We considered some alternatives for adding strong preemption, such as special ``initial'' reactions that would determine at the beginning of a reaction whether a preemption should take place or not, and which might suppress subsequent reactions.
However, we quickly realized that \ac{lf} already has this capability with no additions.
The decision to take a transition is made in target-language code,
and the programmer is free to make that decision early and prevent further computation in that mode.
Thus, the transitions offered by modal reactors already allow to realize both types of preemptive behavior.

\subsection{Transition Types}
\label{sec:transitions}

\autoref{sec:timing-discussion} already discussed why we chose to support delayed transitions at the \ac{lf} level and not immediate transitions.
Some languages, such as \ac{scade} or SCCharts, also offer a \emph{deferred} transition that suppresses the immediate reaction of an entered state, as opposed to non-deferred transitions, that immediately activate the behavior of the target state.
Since modal reactors never execute target states in the same microstep, one might argue that our transitions are always deferred.

Statecharts dialects usually use \emph{priorities} to assign an order to available transitions with the higher priority transition preempting lower ones. Modes in \ac{lf} have exactly the inverse behavior, where the \emph{last} invocation of \code{lf\_set\_mode} determines the target mode.
An implementation that favors the first invocation of \code{lf\_set\_mode} could easily be achieved but we prefer a semantics that conforms with the behavior of ports.

Finally, some statechart dialects also distinguish \emph{deep} and \emph{shallow} history transitions, which differ in whether its affect is propagated downwards in hierarchy or not.
Recalling our transitions semantics from \autoref{sec:semantics} and in \autoref{alg:transition}, it becomes apparent that we currently feature a deep variant.
We briefly considered adding a shallow variant to modal reactors, but did not consider the added complexity that this would have implied as justified.
Again, we would argue that if this behavior would really be desired, it could still be realized at target language level.

\begin{figure}
  \centering

  \subfloat[Example with multiple feeders (reaction and two reactors) to the same output port (\textsf{out}) but separated by modes.]{
    \centering  
    \includegraphics[scale=\lfscale]{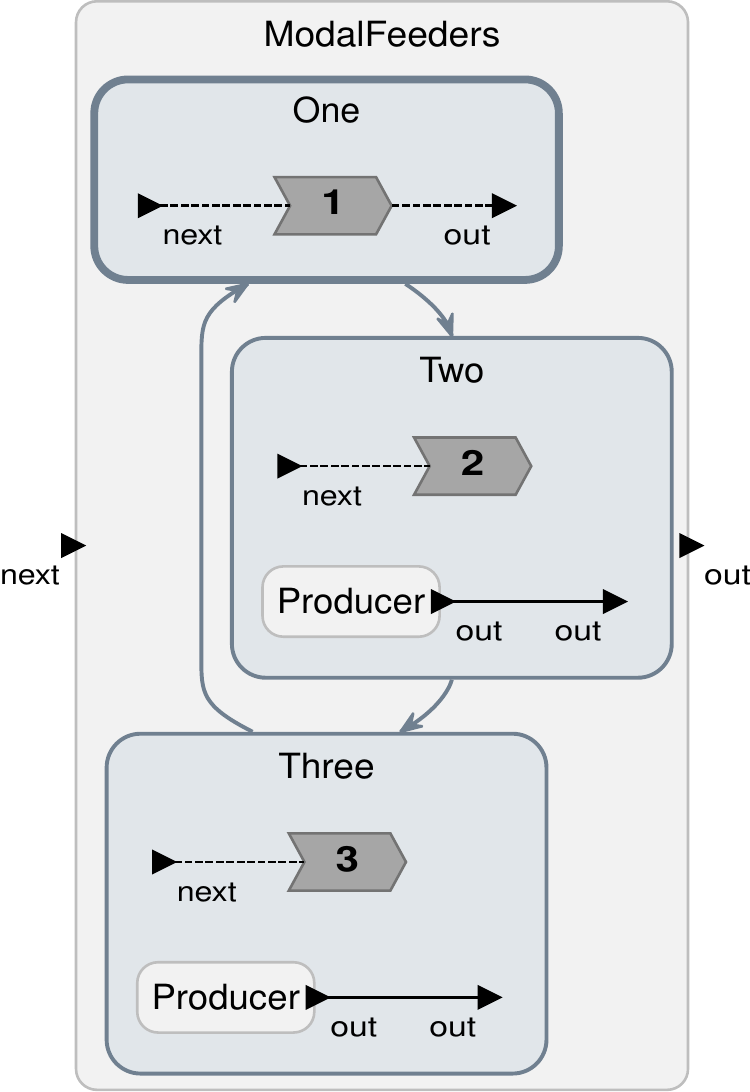}
    \label{fig:modal-feeders}
  }
  \hfill
  \subfloat[Example with a cyclic dependency resolved by the use of modes.]{
    \centering  
    \includegraphics[scale=\lfscale]{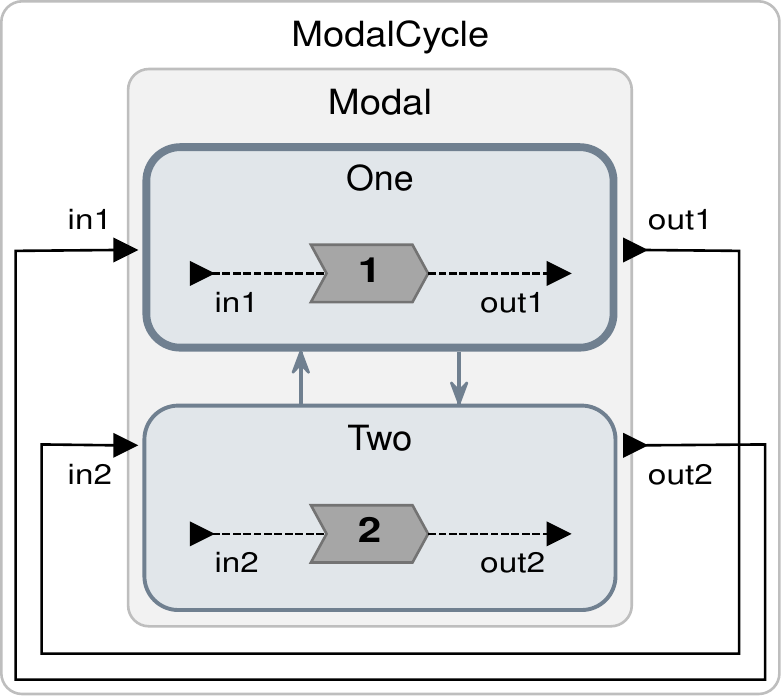}
    \label{fig:modal-cycle}
  }

  \caption{Two examples for \ac{lf} models that can be accepted as determenistic/causal due to the use of modes.}
  \label{fig:mode-benefits}
\end{figure}

\subsection{Modes and Causality}
\label{sec:causality}

As explained, the availability of reset and continue transitions offers a level of control over behavior \wrt time that is not easily achievable using simple in-code workarounds.
The same holds for the possibility to express mutual exclusion structurally with modes.
This lifts certain modeling restrictions imposed under the standard \ac{lf} model of computation.
For example, in the absence of modes, an output port must not be fed by multiple connections to reactors or a mix of reactions and reactors, as this would be a potential source of nondeterminism.
Modes allow such configurations provided that each writer is located in a separate mode to ensure mutual exclusivity between the outputs (see \autoref{fig:modal-feeders} for an example).

The same reasoning applies to causality problems imposed by feedback loops.
The use of modes enables a more fine-grained, less conservative dependency analysis.
For example, the program in \autoref{fig:modal-cycle} would be rejected in the absence of modes due to the following causality loop: \textsf{in1 -- out1 -- in2 -- out2 -- in1}. Mutual exclusivity between the two modes eliminates the causal cycle.

\subsection{Modes as Mutations}
\label{sec:mutations}

As presented here, modes are designed as core language feature of \ac{lf}.
At the onset, we also considered to realize modes based on mutations, which are already part of the formal definition of reactors~\cite{Lohstroh:2019:CyPhy}.
However, while this would be possible to a certain degree, there are methodical considerations that argue against this.
First, mutations are supposed to remodel the topology of a \ac{lf} at runtime, while modes are built around the idea of activity and inactivity and as such always statically present.
Second, the creation and destruction of reactors by mutations at this point does not account for any previous state.
This makes the analogy that a transition is the destruction of the old mode and the creation of the new one rather complicated if a continue behavior is desired.
As proposed here, continue is simply the absence of a reset.

\subsection{Formal Analysis of Modes}
\label{sec:verification}

The integration of modes in the language offers new opportunities for verification and model checking, as more structural information is statically present.
However, a limitation of the current design is the location of transition triggering inside the reaction code.
As other effects declared on reactions, a transition effect to some target mode declared at \ac{lf} level only reflects the potential for such an effect.
The actual invocation depends on the execution of the reaction code.
This in turn depends on the presence of any of the triggers and their runtime value.
Hence, there is no real static estimation for active modes without target code execution.
However, this is true in general for any state-dependent verification in \ac{lf}.
With state variables and ports representing values in the target languages, there is an implicit limitation to target language independent simulation.
Hence, the declaration of potential transition targets as effects is in line with the polyglot approach of \ac{lf} and sufficient for diagram generation and structural analyses.
The extension of verification capabilities including modes is currently considered for the future development of \ac{lf}.

\section{Related Work}
\label{sec:related-work}

We are not aware of other work that aims to create a polyglot modal coordination layer based on a reactor-like foundation.
However, there clearly is significant related work that our proposal builds on.

The synchronous modeling language \ac{scade}~\cite{Colaco:17:SCADE} originally had a dataflow focus, where concurrently operating nodes communicate via streams.
However, from version 6 onwards, it includes state machines as well, based on the mode extension proposed by Cola\c{c}o et al.~\cite{ColacoHP:06:Modes}.
Their work is perhaps closest in spirit to what we propose here, also because they manage to enable modes through a minimal language extension.
Specifically, they introduce modes, quite elegantly, by extending originally boolean clocks, which control execution in Lustre/\ac{scade}, to a richer type that encodes modes.
\ac{scade} represents a standalone language and while it is able to integrate with its target languages and coordinate behavior through signals, it does not embody the rigorous polyglot coordination nature envisioned by \ac{lf}.

More generally, there are many state machine notations, beyond flat \acp{fsm}, that offer feature-rich language constructs.
The most prominent and powerful ones are statecharts by Harel~\cite{harel1987statecharts}, which realize hierarchical state machines and are also part of \ac{uml}.
The statechart dialects that emerged in the context of synchronous languages, such as SyncCharts~\cite{Andre:04:SyncCharts} or SCCharts~\cite{vonHanxledenPLDI14}, are of particular relevance for modes in \ac{lf} as their semantics also involve a clear notion of time and reactions.
There is some work on incorporating physical time into the synchronous model of computation in a way that preserves determinism~\cite{SchulzRosengartenvHMdS+18a,Edwards:20:SparseSynchronous}.
However, typically a ``multiform notion of time'' is based on counting events, such as the passage of 1 msec.

It is also common practice to express statecharts directly in classical programming languages without real language extensions, similar to \autoref{fig:furuta-non-modal}.
Samek describes how to express UML Statecharts in C/C++~\cite{Samek:08:StatechartsC}.
As in UML Statecharts, this approach does not provide deterministic concurrency.
Wagner \etal describe how to implement FSMs in C~\cite{Wagner:06:FSMs}, but these are flat automata without any concurrency.
Moreover, none of them follows a polyglot coordination approach.

Similarly, the Akka framework~\cite{AkkaAction2016} provides means to write actor networks directly in Java, but without consideration of modes or determinism.

Various proposals have been made to augment mainstream programming languages, such as C, with a concept of state or modes.
The Esterel-C Language (ECL)~\cite{Lavagno:99:ECL} is a proposal to extend C by Esterel-like constructs for signal handling and reactive control flow, and from this program the ECL compiler derives an Esterel part and a purely sequential C part.
FairThreads~\cite{Boussinot:06:FairThreads} is another extension of C inspired by Esterel, implemented via native threads, that offers macros to express automata.
Precision Timed C (PRET-C)~\cite{Andalam:10:PRET-C}, which focuses on temporal predictability and assumes a target architecture with specific support for thread scheduling and abort handling, and ForeC~\cite{Yip:16:ForeC}, which targets multi-core architectures, introduce a modal behavior into C programs via pause statements.
Among these synchronous extensions to C, perhaps closest to our proposal are SyncCharts in C, which augment C with a light-weight language extension, realized as C macros, that provides modes based on SyncCharts~\cite{vonHanxleden:09:SC}.
The macros realize states and concurrency by controlling execution with coarse-grain program counters, technically realized with ordinary C labels and computed gotos.
Determinism is achieved by dispatching concurrent threads according to (typically static) priorities.

\section{Conclusion and Outlook}
\label{sec:conclusions-outlook}

Modal reactors enable the coordination of reactive behavior in terms of modes and transitions.
While we capitalize of many existing concepts of \ac{lf} to achieve our objectives of being  lean, polyglot, concurrent, timed, and deterministic, our design carefully adapts these fundamental principles and seamlessly integrates into the existing language, diagrams, and tooling.
Modes are a fundamental concept in real-time systems and how designers think about them.
They go beyond specifying low-level stateful behavior in state machines.
Modal coordination in this context enables the orchestration of complex event processing networks associated with different modes of operation.
Additionally, with mode-local time, it grants a powerful tool for modeling timed-behavior.
Pausing and continuing mode-local behavior is a capability that is otherwise tedious to achieve in \ac{lf}.

While modes are already central to a large family of existing programming and
modeling languages, our approach of building modal abstractions into the polyglot
coordination language \ac{lf} has the advantage of being applicable to a range of target languages at once.
Programmers can still develop the low-level ``business logic'' in any target language supported by \ac{lf}, and may use \ac{lf} solely to express modal aspects in a lean, deterministic manner, with diagramming support, largely irrespective of which other \ac{lf} facilities might be harnessed as well.
Our implementation currently provides modal support for the C and Python targets, demonstrating the versatility of our approach.
While other targets will follow, the C target illustrates the suitability for embedded low-level applications, while Python shows compatibility with a high-level scripting language 
commonly used in Cloud and machine learning applications.
We also successfully used modes in \ac{lf} to control a robot and specify different modes for driving and collision avoidance.
It will be used to teach students the modeling and design of embedded systems.

A key design decision of modal reactors is that we want to preserve the clear separation of coordination language (\ac{lf}) and target language (C, Python, Rust, \ldots).
At the \ac{lf} level, we express the modal structure, modes encapsulating components and transitions.
We go so far as to define which modes are initial, and which reactions may trigger which transitions.
However, the actual run-time behavior, concerning invoking transitions and effects, is handled in the target language via reactions.
Yet, the role of the target language only limited to triggering transitions, while the entire modal infrastructure and behavior is automatically generated.
Due to the separation at the \ac{lf} level, modes are abstractions of actual behavior.
Whether a mode is actually reachable or not is not specified at the \ac{lf} level but depends on the implementation.

While the static aspects of modes already yield a certain amount of structural information that could be used for model checking, a natural question is whether a holistic analysis of both is feasible and worthwhile.
To some extent, this is already taking place during compilation, where for example type compatibility or, in the case of nodes, the consistency of mode transitions is checked.
However, as discussed in \ref{sec:verification}, a possible future work would be to perform model checking on \ac{lf} programs in a ``white-box'' manner that also considers the target language, akin for example to SPIN~\cite{Holzmann:00:CSPIN}.
We also plan to develop tools for verification and model checking of modal reactors.
Another direction is the exploration of modal models coordinating federated \ac{lf} programs.

Finally, we are interested exploring whether modes or alternative (minimal) language extensions can be used to express the behavior trees in \ac{lf}.

\bibliographystyle{IEEEtranS}
\bibliography{../../Refs,../Refs}

\begin{thebibliography}{10}
\providecommand{\url}[1]{#1}
\csname url@samestyle\endcsname
\providecommand{\newblock}{\relax}
\providecommand{\bibinfo}[2]{#2}
\providecommand{\BIBentrySTDinterwordspacing}{\spaceskip=0pt\relax}
\providecommand{\BIBentryALTinterwordstretchfactor}{4}
\providecommand{\BIBentryALTinterwordspacing}{\spaceskip=\fontdimen2\font plus
\BIBentryALTinterwordstretchfactor\fontdimen3\font minus
  \fontdimen4\font\relax}
\providecommand{\BIBforeignlanguage}[2]{{%
\expandafter\ifx\csname l@#1\endcsname\relax
\typeout{** WARNING: IEEEtranS.bst: No hyphenation pattern has been}%
\typeout{** loaded for the language `#1'. Using the pattern for}%
\typeout{** the default language instead.}%
\else
\language=\csname l@#1\endcsname
\fi
#2}}
\providecommand{\BIBdecl}{\relax}
\BIBdecl

\bibitem{Andalam:10:PRET-C}
S.~Andalam, P.~S. Roop, and A.~Girault, ``Predictable multithreading of
  embedded applications using {PRET-C},'' in \emph{Formal Methods and Models
  for Codesign (MEMOCODE)}.\hskip 1em plus 0.5em minus 0.4em\relax IEEE/ACM,
  2010, Conference Proceedings, pp. 159--168.

\bibitem{Andre:04:SyncCharts}
C.~Andr{\'e}, ``Computing {SyncCharts} reactions,'' \emph{Electronic Notes in
  Theoretical Computer Science}, vol.~88, pp. 3--19, Oct. 2004.

\bibitem{1173191}
A.~Benveniste, P.~Caspi, S.~Edwards, N.~Halbwachs, P.~Le~Guernic, and
  R.~de~Simone, ``The synchronous languages 12 years later,'' \emph{Proceedings
  of the IEEE}, vol.~91, no.~1, pp. 64--83, 2003.

\bibitem{Boussinot:06:FairThreads}
F.~Boussinot, ``Fairthreads: mixing cooperative and preemptive threads in
  {C},'' \emph{Concurrency and Computation: Practice and Experience}, vol.~18,
  no.~5, pp. 445--469, Apr. 2006.

\bibitem{ColacoHP:06:Modes}
J.-L. Cola\c{c}o, G.~Hamon, and M.~Pouzet, ``Mixing signals and modes in
  synchronous data-flow systems,'' in \emph{ACM International Conference on
  Embedded Software (EMSOFT'06)}.\hskip 1em plus 0.5em minus 0.4em\relax Seoul,
  South Korea: ACM, Oct. 2006, pp. 73--82.

\bibitem{Colaco:17:SCADE}
J.~Cola{\c{c}}o, B.~Pagano, and M.~Pouzet, ``{SCADE} 6: {A} formal language for
  embedded critical software development (invited paper),'' in \emph{11th
  International Symposium on Theoretical Aspects of Software Engineering
  {TASE}}, Sophia Antipolis, France, Sep. 2017, pp. 1--11.

\bibitem{Conway:63:Coroutines}
M.~E. Conway, ``Design of a separable transition-diagram compiler,''
  \emph{Communications of the ACM}, vol.~6, no.~7, pp. 396--408, 1963.

\bibitem{eventdrivenprog}
F.~Dabek, N.~Zeldovich, F.~Kaashoek, D.~Mazi\`{e}res, and R.~Morris,
  ``Event-driven programming for robust software,'' in \emph{Proceedings of the
  10th Workshop on ACM SIGOPS European Workshop}, ser. EW 10.\hskip 1em plus
  0.5em minus 0.4em\relax New York, NY, USA: Association for Computing
  Machinery, 2002, p. 186–189.

\bibitem{Edwards:20:SparseSynchronous}
S.~A. Edwards and J.~Hui, ``The sparse synchronous model,'' in \emph{2020 Forum
  for Specification and Design Languages (FDL)}, 2020, pp. 1--8.

\bibitem{furuta1992swing}
K.~Furuta, M.~Yamakita, and S.~Kobayashi, ``Swing-up control of inverted
  pendulum using pseudo-state feedback,'' \emph{Proceedings of the Institution
  of Mechanical Engineers, Part I: Journal of Systems and Control Engineering},
  vol. 206, no.~4, pp. 263--269, 1992.

\bibitem{hamon2004operational}
G.~Hamon and J.~Rushby, ``An operational semantics for stateflow,'' in
  \emph{International Conference on Fundamental Approaches to Software
  Engineering}.\hskip 1em plus 0.5em minus 0.4em\relax Springer, 2004, pp.
  229--243.

\bibitem{harel1987statecharts}
D.~Harel, ``Statecharts: A visual formalism for complex systems,''
  \emph{Science of computer programming}, vol.~8, no.~3, pp. 231--274, 1987.

\bibitem{Hewitt:77:Actors}
C.~Hewitt, ``Viewing control structures as patterns of passing messages,''
  \emph{Journal of Artificial Intelligence}, vol.~8, no.~3, pp. 323--363, 1977.

\bibitem{Holzmann:00:CSPIN}
G.~J. Holzmann, ``Logic verification of {ANSI-C} code with {SPIN},'' in
  \emph{Proceedings of the 7th International SPIN Workshop on SPIN Model
  Checking and Software Verification}.\hskip 1em plus 0.5em minus 0.4em\relax
  Berlin, Heidelberg: Springer-Verlag, 2000, p. 131–147.

\bibitem{Lavagno:99:ECL}
L.~Lavagno and E.~Sentovich, ``{ECL}: a specification environment for
  system-level design,'' in \emph{Proc.\ 36th ACM/IEEE Conf.\ on Design
  Automation (DAC'99)}.\hskip 1em plus 0.5em minus 0.4em\relax ACM, 1999, pp.
  511--516.

\bibitem{Lee:06:Threads}
E.~A. Lee, ``The problem with threads,'' \emph{Computer}, vol.~39, no.~5, pp.
  33--42, 2006.

\bibitem{LeeTripakis:10:ModalModels}
E.~A. Lee and S.~Tripakis, ``Modal models in {Ptolemy},'' in \emph{3rd
  International Workshop on Equation-Based Object-Oriented Modeling Languages
  and Tools (EOOLT)}, vol.~47.\hskip 1em plus 0.5em minus 0.4em\relax
  Link\"{o}ping University Electronic Press, Link\"{o}ping University, 2010,
  pp. 11--21.

\bibitem{lee2005causality}
E.~A. Lee, H.~Zheng, and Y.~Zhou, ``Causality interfaces and compositional
  causality analysis,'' \emph{Foundations of Interface Technologies (FIT),
  Satellite to CONCUR, San Francisco, CA}, vol.~2, pp. 402--405, 2005.

\bibitem{liu2002realistic}
J.~Liu, J.~Eker, J.~W. Janneck, and E.~A. Lee, ``Realistic simulations of
  embedded control systems,'' \emph{IFAC Proceedings Volumes}, vol.~35, no.~1,
  pp. 391--396, 2002.

\bibitem{Lohstroh:EECS-2020-235}
\BIBentryALTinterwordspacing
M.~Lohstroh, ``Reactors: A deterministic model of concurrent computation for
  reactive systems,'' Ph.D. dissertation, EECS Department, University of
  California, Berkeley, Dec 2020. [Online]. Available:
  \url{http://www2.eecs.berkeley.edu/Pubs/TechRpts/2020/EECS-2020-235.html}
\BIBentrySTDinterwordspacing

\bibitem{Lohstroh:2019:CyPhy}
M.~Lohstroh, {\'I}.~{\'I}ncer~Romeo, A.~Goens, P.~Derler, J.~Castrillon, E.~A.
  Lee, and A.~Sangiovanni-Vincentelli, ``Reactors: A deterministic model for
  composable reactive systems,'' in \emph{8th International Workshop on
  Model-Based Design of Cyber Physical Systems (CyPhy'19)}, vol. LNCS
  11971.\hskip 1em plus 0.5em minus 0.4em\relax Springer-Verlag, 2019,
  Conference Proceedings, p.~27.

\bibitem{LohstrohEtAl:21:Towards}
M.~Lohstroh, C.~Menard, S.~Bateni, and E.~A. Lee, ``Toward a {Lingua Franca}
  for deterministic concurrent systems,'' \emph{ACM Transactions on Embedded
  Computing Systems (TECS), Special Issue on FDL'19}, vol.~20, no.~4, p.
  Article 36, May 2021.

\bibitem{lohstroh2020language}
M.~Lohstroh, C.~Menard, A.~Schulz-Rosengarten, M.~Weber, J.~Castrillon, and
  E.~A. Lee, ``A language for deterministic coordination across multiple
  timelines,'' in \emph{2020 Forum for Specification and Design Languages
  (FDL)}.\hskip 1em plus 0.5em minus 0.4em\relax IEEE, 2020, pp. 1--8.

\bibitem{10.1145/3170472.3133846}
M.~C. Loring, M.~Marron, and D.~Leijen, ``Semantics of asynchronous
  javascript,'' \emph{SIGPLAN Not.}, vol.~52, no.~11, p. 51–62, oct 2017.

\bibitem{manna1992verifying}
Z.~Manna and A.~Pnueli, ``Verifying hybrid systems,'' in \emph{Hybrid
  Systems}.\hskip 1em plus 0.5em minus 0.4em\relax Springer, 1992, pp. 4--35.

\bibitem{AkkaAction2016}
R.~Roestenburg, R.~Bakker, and R.~Williams, \emph{Akka In Action}.\hskip 1em
  plus 0.5em minus 0.4em\relax Manning Publications Co., 2016.

\bibitem{Samek:08:StatechartsC}
M.~Samek, \emph{Practical UML Statecharts in C/C++Event-Driven Programming for
  Embedded Systems}.\hskip 1em plus 0.5em minus 0.4em\relax Newnes, 2008.

\bibitem{SchulzRosengartenvHMdS+18a}
A.~Schulz-Rosengarten, R.~von Hanxleden, F.~Mallet, R.~de~Simone, and
  J.~Deantoni, ``Time in {SCCharts},'' in \emph{Proc.\ Forum on Specification
  and Design Languages (FDL '18)}, Munich, Germany, September 2018.

\bibitem{SchulzeSvH14}
C.~D. Schulze, M.~Sp{\"o}nemann, and R.~von Hanxleden, ``Drawing layered graphs
  with port constraints,'' \emph{Journal of Visual Languages and Computing,
  Special Issue on Diagram Aesthetics and Layout}, vol.~25, no.~2, pp. 89--106,
  2014.

\bibitem{OOPStroustrup}
B.~Stroustrup, ``What is object-oriented programming?'' \emph{IEEE Software},
  vol.~5, no.~3, pp. 10--20, 1988.

\bibitem{vonHanxleden:09:SC}
R.~von Hanxleden, ``Sync{C}harts in {C}---{A} proposal for light-weight
  deterministic concurrency,'' in \emph{ACM Embedded Software Conference
  (EMSOFT)}, 2009, Conference Proceedings, pp. 11--16.

\bibitem{vonHanxledenPLDI14}
R.~von Hanxleden \emph{et~al.}, ``{SCCharts}: Sequentially constructive
  {Statecharts} for safety-critical applications,'' in \emph{ACM SIGPLAN Conf.
  on Programming Language Design and Implementation}, ser. PLDI '14.\hskip 1em
  plus 0.5em minus 0.4em\relax New York, NY, USA: ACM, 2014, pp. 372--383.

\bibitem{vonHanxledenLF+22}
R.~von Hanxleden, E.~A. Lee, H.~Fuhrmann, A.~Schulz{-}Rosengarten,
  S.~Domr{\"{o}}s, M.~Lohstroh, S.~Bateni, and C.~Menard, ``Pragmatics twelve
  years later: a report on {Lingua Franca},'' in \emph{11th International
  Symposium on Leveraging Applications of Formal Methods, Verification and
  Validation (ISoLA)}, ser. Lecture Notes in Computer Science, vol.
  13702.\hskip 1em plus 0.5em minus 0.4em\relax Rhodes, Greece: Springer,
  October 2022, pp. 60--89.

\bibitem{Wagner:06:FSMs}
F.~Wagner, R.~Schmuki, P.~Wolstenholme, and T.~W. Thomas, \emph{Modeling
  Software with Finite State Machines: A Practical Approach}.\hskip 1em plus
  0.5em minus 0.4em\relax Auerbach Publications, 2006.

\bibitem{Yip:16:ForeC}
E.~Yip, A.~Girault, P.~S. Roop, and M.~Biglari{-}Abhari, ``The {ForeC}
  synchronous deterministic parallel programming language for multicores,'' in
  \emph{10th {IEEE} International Symposium on Embedded Multicore/Many-core
  Systems-on-Chip, {MCSOC} 2016, Lyon, France, September 21-23, 2016}.\hskip
  1em plus 0.5em minus 0.4em\relax {IEEE} Computer Society, 2016, pp. 297--304.

\end{thebibliography}

\end{document}